\documentclass[12pt]{article}
\usepackage{amsmath}
\usepackage{graphicx}
\usepackage{color}
\begin{document}
\baselineskip=18 pt
\begin{center}
{\large{\bf Quantum influence of topological defects on a relativistic scalar particle with Cornell-type potential in  cosmic string space-time with a spacelike dislocation }}
\end{center}

\vspace{.5cm}

\begin{center}
{\bf Faizuddin Ahmed}\footnote{faizuddinahmed15@gmail.com ; faiz4U.enter@rediffmail.com}\\ 
{\bf Ajmal College of Arts and Science, Dhubri-783324, Assam, India}
\end{center}

\vspace{.5cm}

\begin{abstract}

We study the relativistic quantum of scalar particles in the cosmic string space-time with a screw dislocation (torsion) subject to a uniform magnetic field including the magnetic quantum flux in the presence of potential. We solve the Klein-Gordon equation with a Cornell-type scalar potential in the considered framework and obtain the energy eigenvalues and eigenfunctions and analyze a relativistic analogue of the Aharonov-Bohm effect for bound states.

\end{abstract}

{\bf keywords:} Relativistic wave equations, electromagnetic field, Aharonov-Bohm effect, topological defects, scalar potential.

\vspace{0.3cm}


\section{Introduction}

In relativistic quantum mechanics, study of spin-$0$ scalar particles via the Klein-Gordon equation on curved the background with the cosmic string has been of current research interest. Several authors have investigated the physical properties of a series of background with G\"{o}del-type geometries, such as, the relativistic quantum dynamics of a scalar particle \cite{aa1,aa2}, spin-$0$ massive charged particles in the presence of a uniform magnetic field with the cosmic string \cite{aa3}, quantum influence of topological defects \cite{aa4}, linear confinement of a scalar particle with the cosmic string \cite{aa5,aa6,aa7}. Furthermore, the relativistic quantum dynamics of spin-$0$ scalar particles was investigated in \cite{aa8,aa9} and observed the similarity of the energy eigenvalues with the Landau levels \cite{aa10}. The relativistic quantum dynamics of a scalar particle in the presence of a homogeneous magnetic field within the Kaluza-Klein theory was investigated in \cite{aa12}. The relativistic quantum dynamics of spin-$0$ massive charged particles in a four-dimensional curved space-time with the cosmic string was studied in \cite{aa15}. Survey on the Klein-Gordon equation in the G\"{o}del-type space-times was studied in \cite{HS}. Furthermore, spin-$0$ scalar particle \cite{AOP}, Klein-Gordon oscillator \cite{AOP2}, generalized Klein-Gordon oscillator subject to a Coulomb-type scalar potential \cite{GERG}, linear confinement of scalar particle \cite{AOP3}, quantum effects on spin-$0$ charged particles with a Coulomb-type scalar and vector potentials \cite{GERG2} in (1+2)-dimensional rotational symmetry space-time backgrounds have been investigated. In addition, the relativistic quantum dynamics of spin-half particles have also been investigated ({\it e. g.}, \cite{aa1,ERFM,EPaa123,MdM2,EPaa122}). The Dirac equation in $(1+2)$-dimensional rotational symmetry space-time was investigated in \cite{EPJP2}.

The cosmic string space-time in the polar coordinates $(t, r, \phi, z)$ is described by the following line element \cite{ERFM,AV3,MH,AV,AV2,WAH,BL} :
\begin{equation}
ds^2= - dt^2 + dr^2 +\alpha^2 r^2\,d\phi^2 + dz^2,
\label{a1}
\end{equation}
where $\alpha=1-4\mu $ is the topological parameter with $\mu$ being the linear mass density of the cosmic string. In  cosmic string space-time, the parameter $\mu$ assumes values in the interval $ 0 < \mu < 1$ within the general relativity \cite{MOK,CF}. Furthermore, in the cylindrical symmetry we have that $ 0 < r < \infty$, $ 0 \leq \phi \leq  2\pi$ and $-\infty < z < \infty$. Cosmic string may have been produced by phase transitions in the early universe \cite{TWBK} as it is predicted in the extensions of the standard model of particle physics \cite{AV3,MH}. Several authors have studied the relativistic quantum mechanics in the cosmic string space-time ({\it e. g.}, \cite{ERFM,LCNS2,H1,H2,H4,H5,MS,MdM,bb1,bb2,bb4,bb6,bb7,bb8,bb9,bb10}).

Various potentials have been used to investigate the bound state solutions to the relativistic wave-equations. Among them, much attention has given on the Cornell potential. The Cornell potential, which consists of a linear potential plus a Coulomb potential, is a particular case of the quark-antiquark interaction, one more harmonic type term \cite{MKB}. The Coulomb potential is responsible for the interaction at small distances and linear potential leads to the confinement. Recently, the Cornell potential has studied in the ground state of three quarks \cite{CA}. However, this type of potential is worked on spherical symmetry; in cylindrical symmetry, which is our case, this type of potential is known as Cornell-type potential \cite{aa3}. Investigation of the relativistic wave-equations with this type of potential are the Klein-Gordon scalar field in spinning cosmic string space-time \cite{MH2}, relativistic quantum dynamics of scalar particle subject to a uniform magnetic field in cosmic string space-time \cite{ERFM}, spin-$0$ scalar particle in $(1+2)$-dimensional rotational symmetry space-time \cite{MPLA}, Aharonov-Bohm effect for bound states \cite{FA}, quantum effects of confining potential on the Klein-Gordon oscillator \cite{RLLV3}, effects of potential on a position-dependent mass system \cite{EVBL3} etc.. Other investigations with this type of interaction are in \cite{MSC,RLLV4,RLLV2,ALCO}. The Cornell-type potential is given by
\begin{equation}
S=\frac{\eta_c}{r}+\eta_{L}\,r,
\label{cc1}
\end{equation}
where $\eta_c,\eta_L$ are the potential parameters.

In \cite{ERFM}, authors studied the relativistic quantum dynamics of bosonic charged particles in the presence of an external fields in a cosmic string space-time. They solved the Klein-Gordon equation and obtained the relativistic energy eigenvalues and wave-function. In addition, they introduced a Cornell-type scalar potential by modifyong the mass term in the Klein-Gordon equation and obtained the bound states solution of the relativistic quantum system. In \cite{RLLV5}, authors studied a spin-$0$ scalar massive charged particle in the presence of an external fields including a magnetic quantum flux in the space-time with a spacelike dislocation under the influence of linear potential. They solved the Klein-Gordon equation and evaluated the energy eigenvalues and analyze a relativistic analogue of the Aharonov-Bohm effect for bound states. In addition, they introduced a linear scalar potential by modifying the mass term in the Klein-Gordon equation and obtained the bound states solution of the relativistic quantum system. In \cite{RLLV}, authors investigated a spin-$0$ scalar charged particles in the presence of an external fields including a magnetic quantum flux in the space-time with a spacelike dislocation subject to a Coulomb-type potential. They solved the Klein-Gordon equation and evaluated the bound states solution of the relativistic quantum system and analyze a relativistic analogue of the Aharonov-Bohm effect for bound states.

Our main motivation in this work is to investigate a relativistic analogue of the Aharonov-Bohm effect \cite{MP,VBB} for bound states of a relativistic scalar charged particle subject to a homogeneous magnetic field including a magnetic quantum flux in the presence of a Cornell-type potential in the cosmic string space-time with a spacelike dislocation. We solve the Klein-Gordon equation in the considered framework and obtain the relativistic energy eigenvalues and eigenfunctions and analyze the effects on the eigenvalues. In addition, we check the role play by the torsion parameter in this relativistic system and see that the presence of torsion parameter modify the energy levels and break their degeneracy in comparison to the result obtained in the cosmic string space-time case.

\section{Bosonic charged particles : The KG-equation }

In \cite{RAP,VBB10}, examples of topological defects in the space-time associated with torsion are given. We start this section by considering the cosmic string space-time with a spacelike dislocation, whose line element is given by ($x^0=t$, $x^1=r$, $x^2=\phi$, $x^3=z$)
\begin{equation}
ds^2=-dt^2+dr^2+\alpha^2\,r^2\,d\phi^2+(dz+\chi\,d\phi)^2,
\label{b1}
\end{equation}
where $\alpha >0$ is the cosmic string, and $\chi$ is the dislocation (torsion) parameter. For zero torsion parameter, $\chi \rightarrow 0$, the metric (\ref{b1}) reduces to the cosmic string space-time. Furthermore, for $\chi \rightarrow 0$ and $\alpha \rightarrow 1$, the study space-time reduces to Minkowski flat space metric in cylindrical coordinates. Topological defects associated with torsion have investigated in solid state \cite{MOK,HK,CF3,JW,CFil}, quantum scattering \cite{GAM2}, bound states solutions \cite{FA,RLLV5,CF4}, and in relativistic quantum mechanics \cite{RLLV6, RLLV8}.

The metric tensor for the space-time (\ref{b1}) to be 
\begin{equation}
g_{\mu\nu} ({\bf x})=\left (\begin{array}{llll}
-1 & 0 & \quad 0 & 0 \\
0 & 1 & \quad 0 & 0 \\
0 & 0 & \alpha^2\,r^2+\chi^2 & \chi \\
0 & 0 & \quad \chi & 1
\end{array} \right)
\label{b2}
\end{equation}
with its inverse 
\begin{equation}
g^{\mu\nu} ({\bf x})=\left (\begin{array}{llll}
-1 & 0 & \quad 0 & \quad 0 \\
\quad 0 & 1 & \quad 0 & \quad 0 \\
\quad 0 & 0 & \frac{1}{\alpha^2\,r^2} & -\frac{\chi}{\alpha^2\,r^2} \\
\quad 0 & 0 & -\frac{\chi}{\alpha^2\,r^2} & 1+\frac{\chi^2}{\alpha^2\,r^2}
\end{array} \right)
\label{b3}
\end{equation}
The metric has signature $(-,+,+,+)$ and the determinant of the corresponding metric tensor $g_{\mu\nu}$ is
\begin{equation}
det\;g=-\alpha^2\,r^2.
\label{b4}
\end{equation}

The relativistic quantum dynamics of spin-$0$ charged scalar particles of mass $m$ is described by the Klein-Gordon (KG) equation \cite{ERFM,LCNS}
\begin{equation}
\left[\frac{1}{\sqrt{-g}}\,{\sf D}_{\mu} (\sqrt{-g}\,g^{\mu\nu}\,{\sf D}_{\nu})-m^2 \right]\,\Psi=0,
\label{b5}
\end{equation}
where the minimal substitution is defined by
\begin{equation}
{\sf D}_{\mu} \equiv \partial_{\mu}-{\sf i}\,e\,A_{\mu},
\label{b6}
\end{equation}
where $e$ is the electric charge, and $A_{\mu}$ is the electromagnetic four-vector potential by
\begin{equation}
A_{\mu}=(0,\vec{A})\quad,\quad \vec{A}=(0,A_{\phi},0).
\label{b7}
\end{equation}
We choose the angular component of electromagnetic four-vector potential \cite{FA,MSC,RLLV2,ALCO,RLLV5,RLLV,RLLV6,RLLV7}
\begin{equation}
A_{\phi}=-\frac{1}{2}\,\alpha\,B_0\,r^2+\frac{\Phi_B}{2\,\pi}\quad,\quad \vec{B}=\vec{\nabla} \times \vec{A}=-B_0\,\hat{k}.
\label{b8}
\end{equation}
Here $\Phi_B=const.$ is the internal quantum magnetic flux \cite{YA,GAM} through the core of the topological defects \cite{CF6}. It is noteworthy that the Aharonov-Bohm effect \cite{MP,VBB} has been investigated in several branches of physics, such as in, graphene \cite{RJ}, Newtonian theory \cite{MAA}, bound states of massive fermions \cite{VRK},  scattering of dislocated wave fronts \cite{CC}, torsion effects on a relativistic position-dependent mass system \cite{FA,RLLV5}, Kaluza-Klein theory \cite{EVBL3,CF2,CF5,EVBL,aa12,EVBL2,EPaa12}, and non-minimal Lorentz-violating coupling \cite{HB}.

If one introduces a scalar potential by modifying the mass term in the form $m \rightarrow m+ S(r)$ \cite{WG} into the above equation, then we have
\begin{equation}
\left[\frac{1}{\sqrt{-g}}\,{\sf D}_{\mu} (\sqrt{-g}\,g^{\mu\nu}\,{\sf D}_{\nu})-(m+S)^2\right]\,\Psi=0,
\label{b9}
\end{equation}
Several authors have studied the relativistic wave-equations with various kind of potentials such as linear, Coulomb-type, Cornell-type etc. ({\it e. g.},\cite{aa3,aa5,aa7,ERFM,MPLA,FA,RLLV3,RLLV4,RLLV2,RLLV5,RLLV}).

Using the equation (\ref{b1}), Eq. (\ref{b9}) becomes
\begin{eqnarray}
&&[-\frac{\partial^2}{\partial t^2}+\frac{1}{r}\,\frac{\partial}{\partial r}\,(r\,\frac{\partial}{\partial r})+\frac{1}{\alpha^2\,r^2}\,(\frac{\partial}{\partial \phi}-i\,e\,A_{\phi}-\chi\,\frac{\partial}{\partial z})^2+\frac{\partial^2}{\partial z^2}\nonumber\\
&&-(m+S)^2 ]\,\Psi=0.
\label{b10}
\end{eqnarray}

Since the line-element (\ref{b1}) is independent of $t,\phi, z$, it is appropriate to choose the following ansatz for the function $\Psi$
\begin{equation}
\Psi (t, r, \phi, z)={\sf e}^{{\sf i}\,(-E\,t+l\,\phi+k\,z)}\,\psi (r),
\label{b11}
\end{equation}
where $E$ is the energy of charged particle, $l=0,\pm\,1,\pm\,2....\in {\bf Z}$ is the eigenvalues of $z$-component of the angular momentum operator, and $k$ is a constant.

Substituting Eq. (\ref{b11}) into the Eq. (\ref{b10}), we obtain the following radial wave-equation for $\psi (r)$:
\begin{equation}
\left [\frac{d^2}{dr^2}+\frac{1}{r}\,\frac{d}{dr} + E^2-\frac{1}{\alpha^2\,r^2}\,(l-e\,A_{\phi}-k\,\chi)^2-k^2-(m+S)^2 \right]\,\psi (r)=0.
\label{b12}
\end{equation}

Substituting the Eq. (\ref{b8}) and scalar potential (\ref{cc1}) into the Eq. (\ref{b12}), we obtain
\begin{equation}
\left [\frac{d^2}{dr^2}+\frac{1}{r}\,\frac{d}{dr} + \lambda-\omega^2\,r^2-\frac{j^2}{r^2}-\frac{a}{r}-b\,r \right]\,\psi (r)=0,
\label{cc2}
\end{equation}
where
\begin{eqnarray}
&&\lambda=E^2-m^2-k^2-2\,\eta_c\,\eta_L-\frac{2\,m\,\omega_c}{\alpha}\,(l-k\,\chi-\Phi),\nonumber\\
&&\omega=\sqrt{m^2\,\omega^2_{c}+\eta^2_{L}},\nonumber\\
&&j=\sqrt{\frac{(l-k\,\chi-\Phi)^2}{\alpha^2}+\eta^2_{c}},\nonumber\\
&&a=2\,m\,\eta_c,\nonumber\\
&&b=2\,m\,\eta_L, \nonumber\\
&&\omega_c=\frac{e\,B_0}{2\,m}
\label{cc3}
\end{eqnarray}
is called the cyclotron frequency of the particle moving in the magnetic field.

Transforming $x=\sqrt{\omega}\,r$ into the Eq. (\ref{cc2}), we obtain the following equation:
\begin{equation}
\psi ''(x)+\frac{1}{x}\,\psi' (x)+\left[\zeta-x^2-\frac{j^2}{x^2}-\frac{\eta}{x}-\theta\,x \right]\,\psi (x)=0,
\label{cc4}
\end{equation}
where we have defined
\begin{equation}
\zeta=\frac{\lambda}{\omega}\quad,\quad \eta=\frac{a}{\sqrt{\omega}}\quad,\quad \theta=\frac{b}{\omega^{\frac{3}{2}}}.
\label{cc5}
\end{equation}

We now use appropriate boundary conditions to investigate the bound states solutions in this problem. It is require that the wave-functions must be regular both at $x \rightarrow 0$ and $x \rightarrow \infty $. Suppose the possible solution to the Eq. (\ref{cc4}) is
\begin{equation}
\psi (x)=x^{j}\,e^{-\frac{1}{2}\,(\theta+x)\,x}\,H (x),
\label{cc6}
\end{equation}
where $H (x)$ is an unknown function. Substituting the solution (\ref{cc6}) into the Eq. (\ref{cc4}), we obtain
\begin{equation}
H'' (x)+\left[\frac{\gamma}{x}-\theta-2\,x \right]\,H' (x)+\left[-\frac{\beta}{x}+\Theta \right]\,H (x)=0,
\label{cc7}
\end{equation}
where
\begin{eqnarray}
&&\gamma=1+2\,j,\nonumber\\
&&\Theta=\zeta+\frac{\theta^2}{4}-2\,(1+j),\nonumber\\
&&\beta=\eta+\frac{\theta}{2}\,(1+2\,j).
\label{cc8}
\end{eqnarray}
Equation (\ref{cc7}) is the biconfluent Heun's differential equation \cite{aa3,aa5,aa7,aa15,ERFM,MH2,MPLA,FA,RLLV3,EVBL3,RLLV2,RLLV5,RLLV6,RLLV7,EVBL2,EPaa12,AR,SYS} with $H(x)$ is the Heun polynomial function.

The above equation (\ref{cc7}) can be solved by the Frobenius method. Writing the solution as a power series expansion around the origin \cite{GBA}:
\begin{equation}
H (x)=\sum^{\infty}_{i=0}\,c_{i}\,x^{i}.
\label{cc9}
\end{equation}
Substituting the power series solution (\ref{cc9}) into the Eq. (\ref{cc7}), we get the following recurrence relation for the coefficients:
\begin{equation}
c_{n+2}=\frac{1}{(n+2)(n+2+2\,j)}\,\left[\{\beta+\theta\,(n+1)\}\,c_{n+1}-(\Theta-2\,n)\,c_{n} \right].
\label{cc10}
\end{equation}
And the various co-efficients are
\begin{eqnarray}
&&c_1=\left(\frac{\eta}{1+2\,j}+\frac{\theta}{2} \right)\,c_0,\nonumber\\
&&c_2=\frac{1}{4\,(1+j)}\,\left[(\beta+\theta)\,c_{1}-\Theta\,c_{0} \right].
\label{cc11}
\end{eqnarray}

As the function $H (x)$ has a power series expansion around the origin in Eq. (\ref{cc9}), then, the relativistic bound states solution can be achieved by imposing that the power series expansion becomes a polynomial of degree $n$. Through the recurrence relation Eq. (\ref{cc10}), we can see that the power series expansion $H (x)$ becomes a polynomial of degree $n$ by imposing the following two conditions \cite{aa3,aa5,aa7,aa15,ERFM,MH2,MPLA,FA,RLLV3,EVBL3,RLLV2,RLLV5, RLLV6,RLLV7,EVBL2,EPaa12,AVV,JM}
\begin{eqnarray}
\Theta&=&2\,n,\quad (n=1,2,....)\nonumber\\
c_{n+1}&=&0.
\label{cc12}
\end{eqnarray}

By analyzing the first condition $\Theta=2\,n$, we get second degree equation of the energy eigenvalues $E_{n,l}$:
\begin{eqnarray}
E^2_{n,l}&=&\frac{2\,m\,\omega_c}{\alpha}\,(l-k\,\chi-\Phi)+2\,\omega\,(n+1+\sqrt{\frac{(l-k\,\chi-\Phi)^2}{\alpha^2}+\eta^2_{c}})\nonumber\\
&&-\frac{m^2\,\eta^2_{L}}{\omega^2}+m^2+k^2+2\,\eta_c\,\eta_L.
\label{cc13}
\end{eqnarray}

The wave-functions is given by
\begin{equation}
\psi_{n,l} (x)=x^{\sqrt{\frac{(l-\Phi-k\,\chi)^2}{\alpha^2}+\eta^2_{c}}}\, {\sf e}^{-\frac{1}{2}\,\left(\frac{2\,m\,\eta_L}{\omega^{\frac{3}{2}}}+x \right)\,x}\,H (x).
\label{cc15}
\end{equation}

Note that the Eq. (\ref{cc13}) does not represent the general expression for eigenvalues problem. One can obtain the individual energy eigenvalues one by one, that is, $E_1$, $E_2$, $E_3$ by imposing the additional recurrence condition $c_{n+1}=0$ on the eigenvalue. The solution with Heun's Equation makes it possible to obtain the eigenvalues one by one as done in \cite{aa3,aa5,aa7,aa15,ERFM,MH2,MPLA,FA,RLLV3,EVBL3,RLLV2,RLLV5,RLLV6,RLLV7,EVBL2,EPaa12,AVV,JM} but not explicitly in the general form by all eigenvalues $n$. With the aim of obtaining the energy levels of the stationary states, let us discuss a particular case of $n=1$. This means that we want to construct a polynomial of first degree to $H (x)$. With $n=1$, we have $\Theta=2$ and $c_2=0$ which implies from Eq. (\ref{cc11})
\begin{eqnarray}
&&c_1=\frac{2}{(\beta+\theta)}\,c_0 \Rightarrow \frac{\eta}{1+2\,j}+\frac{\theta}{2}=\frac{2}{(\beta+\theta)}\nonumber\\\Rightarrow 
&&\omega^3_{1,l}-\frac{a^2}{2\,(1+2\,j)}\,\omega^2_{1,l}-a\,b\,(\frac{1+j}{1+2\,j})\,\omega_{1,l}-\frac{b^2}{8}\,(3+2\,j)=0.
\label{cc16}
\end{eqnarray}
a constraint on the physical parameter $\omega_{1,l}$. The relation given in Eq. (\ref{cc16}) gives the possible values of the parameter $\omega_{1,l}$ that permit us to construct first degree polynomial to H(x) for $n=1$. Note that its values changes for each quantum number $n$ and $l$, so we have labeled $\omega \rightarrow \omega_{n,l}$. Besides, since this parameter is determined by the frequency or the magnetic field $B_0$, hence, the magnetic field $B_0^{1,l}$ is so adjusted that the Eq. (\ref{cc16}) can be satisfied and the first degree polynomial to $H (x)$ can be achieved, where we have simplified our notation by labeling:
\begin{equation}
\omega_{c\,1,l}=\frac{1}{m}\sqrt{\omega^2_{1,l}-\eta^2_{L}} \leftrightarrow B_{0}^{1,l}=\frac{2}{e}\,\sqrt{\omega^2_{1,l}-\eta^2_{L}}.
\label{ssee}
\end{equation}
It is noteworthy that a third-degree algebraic equation (\ref{cc16}) has at least one real solution and it is exactly this solution that gives us the allowed values of the magnetic field for the lowest state of the system, which we do not write because its expression is very long. We can note, from Eq. (\ref{ssee}) that the possible values of the magnetic field depend on the quantum numbers and the potential parameter. In addition, for each relativistic energy levels, we have different relation of the magnetic field associated to the Cornell-type potential and quantum numbers of the system $\{l, n \}$. For this reason, we have labeled the parameters $\omega$, $\omega_c$ and $B_0$ in Eqs. (\ref{cc16}) and (\ref{ssee}).

Therefore, the ground state energy levels for $n=1$ is given by
\begin{eqnarray}
E_{1,l}&=&\pm\{\frac{2\,m\,\omega_{c\,1,l}}{\alpha}\,(l-k\,\chi-\Phi)+2\,m\,\omega_{1,l}\,(2+\sqrt{\frac{(l-k\,\chi-\Phi)^2}{\alpha^2}+\eta^2_{c}})\nonumber\\
&&-\frac{m^2\,\eta^2_L}{\omega^2_{1,l}}+m^2+k^2+2\,\eta_c\,\eta_L\}^{\frac{1}{2}},
\label{cc17}
\end{eqnarray}
Then, by substituting the real solution $\omega_{1,l}$ from Eq. (\ref{cc16}) into the Eq. (\ref{cc17}) it is possible to obtain the allowed values of the relativistic energy levels for the radial mode $n=1$ of a position dependent mass system. We can see that the lowest energy state is defined by the real solution of the algebraic equation Eq. (\ref{cc16}) plus the expression given in Eq. (\ref{cc17}) for the radial mode $n=1$, instead of $n=0$. This effect arises due to the presence of Cornell-type potential in the system. Note that, it is necessary physically that the lowest energy state is $n=1$ and not $n=0$, otherwise the opposite would imply that $c_1=0$, which requires that the rest mass of the scalar particle be zero that is contrary to the proposal of this investigation.

For $\Phi_B \neq 0$ and $\chi \neq 0$, we can observe in Eq. (\ref{cc17}) there exists an effective angular momentum $l \rightarrow l'=\frac{1}{\alpha}\,(l-\Phi-k\,\chi)$. Thus the relativistic energy levels depend on the geometric phase \cite{YA,GAM} as well as torsion parameter. This dependence of the energy levels on the geometric quantum phase gives rise to the well-known effect called as the Aharonov-Bohm effect for bound states \cite{FA,EVBL3,RLLV5,MP,VBB,CF2, CF5,EVBL,aa12,EVBL2,EPaa12}. Besides, we have that $E_{n, l} (\Phi_B+\Phi_0)=E_{n, l\mp \tau} (\Phi_B)$ where, $\Phi_0=\pm\,\frac{2\,\pi}{e}\,\tau$ with $\tau=0,1,2...$, which means the relativistic energy eigenvalues (\ref{cc17}) is a periodic function of the Aharonov-Bohm geometric quantum phase.

The ground state wave function for $n=1$ is given by
\begin{equation}
\psi_{1,l}=x^{\sqrt{\frac{(l-\Phi-k\,\chi)^2}{\alpha^2}+\eta^2_{c}}}\,{\sf e}^{-\frac{1}{2}\,\left(\frac{2\,m\,\eta_L}{\omega_{1,l}^{\frac{3}{2}}}+x \right)\,x} \times (c_0+c_1\,x),
\label{cc18}
\end{equation}
where
\begin{equation}
c_1=\frac{1}{\sqrt{\omega_{1,l}}}\,\left[\frac{2\,m\,\eta_c}{1+2\,j}+\frac{m\,\eta_L}{\omega_{1,l}} \right]\,c_0.
\label{cc19}
\end{equation}

\vspace{0.3cm}
\subsection{Interactions with linear scalar potential}
\vspace{0.3cm}

We discuss a case corresponds to $\eta_c \rightarrow 0$, that is, only a linear scalar potential in the considered relativistic quantum systems. Therefore, the radial wave-equation (\ref{cc2}) becomes
we obtain
\begin{equation}
\psi'' (r)+\frac{1}{r}\,\psi' (r)+\left[\tilde{\lambda}-\omega^2\,r^2-\frac{j^2}{r^2}-b\,r \right]\,\psi (r)=0,
\label{c1}
\end{equation}
where $\tilde{\lambda}=E^2-m^2-k^2-\frac{2\,m\,\omega_c}{\alpha}\,(l-k\,\chi-\Phi)$.

By changing the variable $x=\sqrt{\omega}\,r$, Eq. (\ref{c1}) becomes
\begin{equation}
\psi'' (x)+\frac{1}{x}\,\psi' (x)+\left[\frac{\tilde{\lambda}}{\omega}-x^2-\frac{j^2}{x^2}-\theta\,r \right]\,\psi (x)=0,
\label{c3}
\end{equation}
Substituting the solution Eq. (\ref{cc6}) into the Eq. (\ref{c3}), we obtain 
\begin{equation}
H'' (x)+\left[\frac{\gamma}{x}-\theta-2\,x \right]\,H' (x)+\left[-\frac{\tilde{\beta}}{x}+\tilde{\Theta} \right]\,H (x)=0,
\label{c4}
\end{equation}
where
\begin{eqnarray}
&&\gamma=1+2\,j,\nonumber\\
&&\tilde{\Theta}=\frac{\tilde{\lambda}}{\omega}+\frac{\theta^2}{4}-2\,(1+j),\nonumber\\
&&\tilde{\beta}=\frac{\theta}{2}\,(1+2\,j).
\label{c5}
\end{eqnarray}
Equation (\ref{c4}) is the biconfluent Heun's differential equation \cite{aa3,aa5,aa7,aa15,ERFM,MH2,MPLA,FA, RLLV3,EVBL3,RLLV2,RLLV5,RLLV6,RLLV7,EVBL2,EPaa12,AR,SYS} with $H(x)$ is the Heun polynomials function.

Substituting the power series solution (\ref{cc9}) into the Eq. (\ref{c4}), we get the following recurrence relation for the coefficients:
\begin{equation}
c_{n+2}=\frac{1}{(n+2)(n+2+2\,j)}\,\left[\{\tilde{\beta}+\theta\,(n+1)\}\,c_{n+1}-(\tilde{\Theta}-2\,n)\,c_{n} \right].
\label{ff1}
\end{equation}
And the various co-efficients are
\begin{eqnarray}
&&c_1=\frac{\theta}{2}\,c_0,\nonumber\\
&&c_2=\frac{1}{4\,(1+j)}\,\left[(\tilde{\beta}+\theta)\,c_{1}-\tilde{\Theta}\,c_{0} \right].
\label{ff2}
\end{eqnarray}

The power series expansion $H (x)$ becomes a polynomial of degree $n$ by imposing the following two conditions \cite{aa3,aa5,aa7,aa15,ERFM,MH2,MPLA,FA,RLLV3,EVBL3,RLLV2,RLLV5,RLLV6,RLLV7,EVBL2,EPaa12,AVV,JM}
\begin{eqnarray}
\tilde{\Theta}&=&2\,n,\quad (n=1,2,....)\nonumber\\
c_{n+1}&=&0.
\label{ff3}
\end{eqnarray}

By analyzing the first condition, we obtain the following energy eigenvalues expression $E_{n,l}$:
\begin{eqnarray}
E^2_{n,l}&=&m^2+k^2+\frac{2\,m\,\omega_c}{\alpha}\,(l-k\,\chi-\Phi)+2\,\omega\,(n+1+\frac{|l-k\,\chi-\Phi|}{\alpha})\nonumber\\
&&-\frac{m^2\,\eta^2_{L}}{\omega^2},
\label{c2}
\end{eqnarray}
where $n=1,2,...$.

The ground state energy levels associated with the radial mode $n=1$ is given by
\begin{eqnarray}
E^2_{1,l}&=&m^2+k^2+\frac{2\,m\,\omega_{c\,1,l}}{\alpha}\,(l-k\,\chi-\Phi)+2\,\omega_{1,l}\,(n+1+\frac{|l-k\,\chi-\Phi|}{\alpha})\nonumber\\
&&-\frac{m^2\,\eta^2_{L}}{\omega^2_{1,l}},
\label{c6}
\end{eqnarray}
where by using Eq. (\ref{ff3}) for $n=1$, we obtain the following constraint
\begin{eqnarray}
&&\omega_{1,l}=\left[ m^2\,\eta^2_{L}\,(3+2\,j) \right]^{\frac{1}{3}},\nonumber\\
&&\omega_{c\,1,l}=\frac{1}{m}\,\sqrt{\left[m^2\,\eta^2_{L}\,(3+2\,j) \right]^{\frac{2}{3}}-\eta^2_{L}},\nonumber\\
&&j=\sqrt{\frac{(l-k\,\chi-\Phi)^2}{\alpha^2}+\eta^2_{c}}.
\label{c7}
\end{eqnarray}

Equation (\ref{c6}) is the ground state energy levels associated with the radial mode $n=1$ of a relativistic scalar charged particle in the presence of an external uniform magnetic field including a magnetic quantum flux in the cosmic string space-time with a spacelike dislocation. For $\alpha \rightarrow 1$, the energy eigenvalues Eq. (\ref{c2}) reduce to the result obtained in \cite{RLLV5}. We can see that the presence of cosmic string parameter ($\alpha$) shifts the energy levels in comparison to those in \cite{RLLV5}. Thus, by comparing the energy eigenvalues Eq. (\ref{cc17}) with Eq. (\ref{c6}), we have the presence of an extra linear potential modifies the relativistic spectrum of energy.

\vspace{0.3cm}
\subsection{Interactions with Coulomb-type potential}
\vspace{0.3cm}

We discuss another case corresponds to $\eta_L \rightarrow 0$, that is, only Coulomb-type scalar potential in the considered relativistic quantum systems. Therefore, the radial wave-equation (\ref{cc2}) becomes
we obtain
\begin{equation}
\psi'' (r)+\frac{1}{r}\,\psi' (r)+\left[\tilde{\lambda}-m^2\,\omega^2_{c}\,r^2-\frac{j^2}{r^2}-\frac{a}{r} \right]\,\psi (r)=0,
\label{d1}
\end{equation}
where $\tilde{\lambda}$ is given earlier.

Let us define $x=\sqrt{m\,\omega_c}\,r$, then Eq. (\ref{d1}) becomes
\begin{equation}
\psi'' (x)+\frac{1}{x}\,\psi' (x)+\left[\tilde{\lambda}-x^2-\frac{j^2}{x^2}-\frac{\delta}{r} \right]\,\psi (x)=0,
\label{d2}
\end{equation}
where $\delta=\frac{a}{\sqrt{m\,\omega_c}}$. By imposing that $\psi (x) \rightarrow 0$ when $x \rightarrow 0$ and $x \rightarrow \infty$, we have
\begin{equation}
\psi (x)=x^{j}\,e^{-\frac{x^2}{2}}\,H (x).
\label{d3}
\end{equation}
By substituting Eq. (\ref{d3}) into the Eq. (\ref{d2}), we obtain the following equation for H(x):
\begin{equation}
H'' (x)+\left[\frac{1+2\,j}{x}-2\,x \right]\,H' (x)+\left[\frac{\tilde{\lambda}}{m\,\omega_c}-2-2\,j-\frac{\delta}{x} \right]\,H (x)=0.
\label{d4}
\end{equation}
Equation (\ref{d4}) is the biconfluent Heun's differential equation \cite{aa3,aa5,aa7,aa15,ERFM,MH2,MPLA,FA,RLLV3,EVBL3,RLLV2,RLLV5,RLLV6,RLLV7,EVBL2,EPaa12,AR,SYS} with $H(x)$ is the Heun polynomials function.

Substituting the power series solution (\ref{cc9}) into the Eq. (\ref{d4}), we get the following recurrence relation for the coefficients:
\begin{equation}
c_{n+2}=\frac{1}{(n+2)(n+2+2\,j)}\,\left[ \delta\,c_{n+1}-(\frac{\tilde{\lambda}}{m\,\omega_c}-2-2\,j-2\,n)\,c_{n} \right].
\label{ff11}
\end{equation}
And the various co-efficients are
\begin{eqnarray}
&&c_1=\frac{\delta}{1+2\,j}\,c_0,\nonumber\\
&&c_2=\frac{1}{4\,(1+j)}\,\left[\delta\,\,c_{1}-(\frac{\tilde{\lambda}}{m\,\omega_c}-2-2\,j)\,c_{0} \right].
\label{ff22}
\end{eqnarray}

The power series expansion $H (x)$ becomes a polynomial of degree $n$ by imposing the following two conditions \cite{aa3,aa5,aa7,aa15,ERFM,MH2,MPLA,FA,RLLV3,EVBL3,RLLV2,RLLV5,RLLV6,RLLV7,EVBL2,EPaa12,AVV,JM}
\begin{eqnarray}
\frac{\tilde{\lambda}}{m\,\omega_c}-2-2\,j&=&2\,n,\quad (n=1,2,....)\nonumber\\
c_{n+1}&=&0.
\label{ff33}
\end{eqnarray}

By analyzing the first condition, we obtain the following equation of eigenvalues $E_{n,l}$:
\begin{equation}
E_{n,l}=\pm\,\sqrt{m^2+k^2+2\,m\,\omega_{c}\,\left[n+1+j+\frac{1}{\alpha}\,(l-k\,\chi-\Phi) \right]} \quad (n=1,2,....).
\label{d5}
\end{equation}

For $n=1$, the ground state energy levels is given by
\begin{equation}
E_{1,l}=\pm\,\sqrt{m^2+k^2+2\,m\,\omega_{c\,1,l}\,\left[2+j+\frac{1}{\alpha}\,(l-k\,\chi-\Phi) \right]},
\label{d6}
\end{equation}
where by using Eq. (\ref{ff33}) for $n=1$, we obtain the following constraint
\begin{equation}
\omega_{c\,1,l}=\frac{2\,m\,\eta^2_{c}}{1+2\,j} \leftrightarrow B^{1,l}_0=\frac{4\,m^2\,\eta^2_{c}}{e\,(1+2\,j)}.
\label{d7}
\end{equation}
Here the magnetic field $B^{1,l}_0$ is so adjusted that the Eq. (\ref{d7}) can be satisfied and a polynomial of first degree to $ H (x)$ can be achieved.

Equation (\ref{d6}) is the energy levels associated with the radial mode $n=1$ of a relativistic scalar charged particle in the presence of an external uniform magnetic field including a magnetic quantum flux in the cosmic string space-time with a spacelike dislocation. For $\alpha \rightarrow 1$, the energy levels Eq. (\ref{d6}) reduces to the result obtained in \cite{RLLV}. Thus, we can see that the presence of the cosmic string parameter ($\alpha$) shifts the energy levels in comparison to those in \cite{RLLV}. Thus, by comparing the energy eigenvalue expression Eq. (\ref{cc17}) with Eq. (\ref{d6}), we can see that the presence of an extra Coulomb-type potential modifies the relativistic energy spectrum of the system.

\vspace{0.3cm}
\subsection{Without torsion parameter}
\vspace{0.3cm}

We discuss here zero torsion parameter, $\chi \rightarrow 0$, in the considered relativistic quantum systems. 

Therefore, the radial wave-equation (\ref{cc2}) becomes
\begin{equation}
\left [\frac{d^2}{dr^2}+\frac{1}{r}\,\frac{d}{dr} + \lambda_0-\omega^2\,r^2-\frac{j^2_{0}}{r^2}-\frac{a}{r}-b\,r \right]\,\psi (r)=0,
\label{ss1}
\end{equation}
where
\begin{eqnarray}
&&\lambda_0=E^2-k^2-m^2-\frac{2\,m\,\omega_c}{\alpha}\,(l-\Phi)-2\,\eta_c\,\eta_L,\nonumber\\
&&j_0=\sqrt{\frac{(l-\Phi)^2}{\alpha^2}+\eta^2_{c}}.
\label{ss2}
\end{eqnarray}

Transforming a new variable $x=\sqrt{\omega}\,r$ into the Eq. (\ref{ss1}), we obtain
\begin{equation}
\left [\frac{d^2}{dx^2}+\frac{1}{x}\,\frac{d}{dx} + \frac{\lambda_0}{\omega}-x^2-\frac{j^2_{0}}{x^2}-\frac{\eta}{x}-\theta\,x \right]\,\psi (r)=0,
\label{ss3}
\end{equation}

Suppose the possible solution to the Eq. (\ref{ss3}) is
\begin{equation}
\psi (x)=x^{j_0}\,e^{-\frac{1}{2}\,(\theta+x)\,x}\,H (x).
\label{ss4}
\end{equation}
Substituting the solution (\ref{cc6}) into the Eq. (\ref{cc4}), we obtain
\begin{eqnarray}
&&H'' (x)+\left[\frac{1+2\,j_0}{x}-\theta-2\,x \right]\,H' (x)\nonumber\\
&+&\left[-\frac{\beta_0}{x}+\frac{\lambda_0}{\omega}+\frac{\theta^2}{4}-2-2\,j_0 \right]\,H (x)=0,
\label{ss5}
\end{eqnarray}
where $\beta_0=\eta+\frac{\theta}{2}\,(1+2\,j_0)$. 

Equation (\ref{ss5}) is the biconfluent Heun's differential equation \cite{aa3,aa5,aa7,aa15,ERFM,MH2,MPLA,FA,RLLV3,EVBL3,RLLV2,RLLV5,RLLV6,RLLV7,EVBL2,EPaa12,AR,SYS} with $H(x)$ is the Heun polynomial function.

Substituting the power series solution (\ref{cc9}) into the Eq. (\ref{ss5}), we get the following recurrence relation for the coefficients:
\begin{eqnarray}
c_{n+2}&=&\frac{1}{(n+2)(n+2+2\,j_0)}\,[\{\eta+\theta\,(n+j_0+\frac{3}{2})\}\,c_{n+1}\nonumber\\
&&-(\frac{\lambda_0}{\omega}+\frac{\theta^2}{4}-2-2\,j_0-2\,n)\,c_n].
\label{ss6}
\end{eqnarray}
And the various co-efficients are
\begin{eqnarray}
&&c_1=\left(\frac{\eta}{1+2\,j_0}+\frac{\theta}{2} \right)\,c_0,\nonumber\\
&&c_2=\frac{1}{4\,(1+j)}\,\left[\{\eta+\theta\,(j_0+\frac{3}{2})\}\,c_{1}-(\frac{\lambda_0}{\omega}+\frac{\theta^2}{4}-2-2\,j_0)\,c_{0} \right].
\label{ss7}
\end{eqnarray}

The power series expansion $H (x)$ becomes a polynomial of degree $n$ by imposing the following two conditions \cite{aa3,aa5,aa7,aa15,ERFM,MH2,MPLA,FA,RLLV3,EVBL3,RLLV2,RLLV5, RLLV6,RLLV7,EVBL2,EPaa12,AVV,JM}
\begin{eqnarray}
\frac{\lambda_0}{\omega}+\frac{\theta^2}{4}-2-2\,j_0&=&2\,n\quad (n=1,2,...)\nonumber\\
c_{n+1}&=&0.
\label{ss8}
\end{eqnarray}

By analyzing the first condition. we obtain the following second degree energy eigenvalues expression $E_{n,l}$:
\begin{eqnarray}
E^2_{n,l}&=&m^2+k^2+\frac{2\,m\,\omega_c\,(l-\Phi)}{\alpha}+2\,m\,\omega\,\left(n+1+\sqrt{\frac{(l-\Phi)^2}{\alpha^2}+\eta^2_{c}} \right)\nonumber\\
&&+2\,\eta_c\,\eta_L-\frac{m^2\,\eta^2_{L}}{\omega^2}.
\label{ss9}
\end{eqnarray}
Equation (\ref{ss9}) is the energy eigenvalues of a relativistic scalar particle with an external uniform magnetic field including a magnetic quantum flux in the cosmic string space-time subject to Cornell-type scalar potential. For zero magnetic quantum flux, $\Phi_B \rightarrow 0$, the energy eigenvalues (\ref{ss9}) is consistent with the result in \cite{ERFM}. Thus, we can see that the energy eigenvalue expression Eq. (\ref{cc13}) get modify in comparison to those in \cite{ERFM} due to the presence of the magnetic quantum flux $\Phi_B$ as well the torsion parameter $\chi$ which break the degeneracy of the energy spectrum.

To obtain the individual energy levels, we impose the additional recurrence condition $c_{n+1}=0$. For example, $n=1$, we have from (\ref{ss7})
\begin{equation} 
\omega^3_{1,l}-\frac{a^2}{2\,(1+2\,j_0)}\,\omega^2_{1,l}-a\,b\,(\frac{1+j_0}{1+2\,j_0})\,\omega_{1,l}-\frac{b^2}{8}\,(3+2\,j_0)=0,
\label{ss10}
\end{equation}
a constraint on the physical parameter $\omega_{1,l}$. The relation given in Eq. (\ref{ss10}) gives the possible values of the parameter $\omega_{1,l}$ that permit us to construct first degree polynomial to $H (x)$ for $n=1$. Note that its values changes for each quantum number $n$ and $l$, so we have labeled $\omega \rightarrow \omega_{n,l}$. Besides, since this parameter is determined by the frequency or the magnetic field $B_0$, hence, the magnetic field $B_0^{1,l}$ is so adjusted that the Eq. (\ref{ss10}) can be satisfied and the first degree polynomial to $H (x)$ can be achieved, where we have simplified our notation by labeling:
\begin{equation}
\omega_{c\,1,l}=\frac{1}{m}\sqrt{\omega^2_{1,l}-\eta^2_{L}} \leftrightarrow B_{0}^{1,l}=\frac{2}{e}\,\sqrt{\omega^2_{1,l}-\eta^2_{L}}.
\label{ss14}
\end{equation}
Note that the equation (\ref{ss10}) has at least one real solution and it is exactly this solution that gives us the allowed values of the magnetic field for the lowest state of the system, which we do not write because its expression is very long. We can note, from Eq. (\ref{ss14}) that the possible values of the magnetic field depend on the quantum numbers and the potential parameter.

The ground state energy levels for $n=1$ is
\begin{eqnarray}
E_{1,l}&=&\pm\{m^2+k^2+2\,\eta_c\,\eta_L+\frac{2\,m\,\omega_{c\,1,l}}{\alpha}\,(l-\Phi)\nonumber\\
&&+2\,m\,\omega_{1,l}\,\left(2+\sqrt{\frac{(l-\Phi)^2}{\alpha^2}+\eta^2_{c}} \right)-\frac{m^2\,\eta^2_{L}}{\omega^2_{1,l}}\}^{\frac{1}{2}},
\label{ss11}
\end{eqnarray}
Then, by substituting the real solution $\omega_{1,l}$ from Eq. (\ref{ss10}) into the Eq. (\ref{ss11}) it is possible to obtain the allowed values of the relativistic energy levels for the radial mode $n=1$ of a position dependent mass system. We can see that the lowest energy state is defined by the real solution of the algebraic equation Eq. (\ref{ss10}) plus the expression given in Eq. (\ref{ss11}) for the radial mode $n=1$, instead of $n=0$. This effect arises due to the presence of Cornell-type potential in the system.

The ground state wave function is
\begin{equation}
\psi_{1,l}=x^{\sqrt{\frac{(l-\Phi)^2}{\alpha^2}+\eta^2_{c}}}\,{\sf e}^{-\frac{1}{2}\,\left(\frac{2\,m\,\eta_L}{\omega_{1,l}^{\frac{3}{2}}}+x \right)\,x}\,(c_0+c_1\,x),
\label{ss12}
\end{equation}
where
\begin{equation}
c_1=\frac{1}{\omega^{\frac{1}{2}}_{1,l}}\,\left[\frac{2\,m\,\eta_c}{1+2\,j_{0}}+\frac{m\,\eta_L}{\omega_{1,l}} \right]\,c_0.
\label{ss13}
\end{equation}

In {\it sub-section 2.1--2.3}, we can see that the relativistic energy eigenvalues depend on the geometric quantum phase \cite{YA,GAM}. This dependence of the energy eigenvalues on the geometric quantum phase gives rise to a relativistic analogue of the Aharonov-Bohm effect for bound states \cite{FA,EVBL3,RLLV5,MP,VBB,CF2,CF5,EVBL,aa12,EVBL2,EPaa12}. Besides, we have that $E_{n, l} (\Phi_B+\Phi_0)=E_{n, l\mp \tau} (\Phi_B)$ where, $\Phi_0=\pm\,\frac{2\,\pi}{e}\,\tau$ with $\tau=0,1,2...$.

\section{Conclusions}

In \cite{ERFM}, authors studied the relativistic quantum dynamics of bosonic charged particle in the presence of an external fields in the cosmic string space-time subject to a Cornell-type potential. In \cite{RLLV5}, authors studied a scalar field in the presence of an external fields including a magnetic quantum flux in the space-time with a spacelike dislocation subject to a linear potential. In \cite{RLLV}, authors investigated a spin-$0$ massive charged particle in the presence of an external fields including a magnetic quantum flux in the space-time with a spacelike dislocation subject to a Coulomb-type potential.

In this paper, we have investigated quantum effects of torsion and topological defects that stems from a space-time with a spacelike dislocation under the influence of a Cornell-type potential in the relativistic quantum system. By solving the Klein-Gordon equation subject to a uniform magnetic field including a magnetic flux in the presence of a Cornell-type potential, we have obtained the energy eigenvalues Eq. (\ref{cc13}) and corresponding eigenfunctions Eq. (\ref{cc15}). By imposing the additional recurrence condition $c_{n+1}=0$, we have obtained the individual energy levels and corresponding wave-function, as for example, $n=1$ and others are in the same way. The presence of the torsion parameter modify the energy levels and break their degeneracy. We have discussed three cases ({\it sub-section 2.1}) for zero Coulomb potential, $\eta_c \rightarrow 0$, ({\it sub-section 2.2}) zero linear potential, $\eta_L \rightarrow 0$, and ({\it sub-section 2.3}) zero torsion parameter $\chi \rightarrow 0$ in the considered relativistic quantum system. In {\it sub-section 2.1--2.2}, we have seen that for $\alpha \rightarrow 1$, the energy eigenvalues Eq. (\ref{c2}) and Eq. (\ref{d6}) are consistent with those results obtained in \cite{RLLV5} and \cite{RLLV}, respectively. Thus, the presence of the cosmic string parameter shifts the energy levels in comparison to those obtained in \cite{RLLV5,RLLV}. Furthermore, by comparing the energy eigenvalues Eq. (\ref{cc13}) with those results obtained in \cite{RLLV5} and \cite{RLLV}, we have seen that the presence of an extra potential term as well the cosmic string parameter modifies the energy eigenvalues. In {\it sub-section 2.3}, for zero zero magnetic quantum flux, $\Phi_B \rightarrow 0$, we have seen that the energy eigenvalues Eq. (\ref{ss9}) is consistent with those result obtained in \cite{ERFM}. Hence, the relativistic energy eigenvalues Eq. (\ref{ss9}) is the extended result in comparison to those in \cite{ERFM} due to the presence of a magnetic quantum flux $\Phi_B$. In addition, by comparing the energy eigenvalues Eq. (\ref{cc13}) with the result obtained in \cite{ERFM}, we have seen that the presence of the torsion parameter $\chi$ as well as the magnetic quantum flux $\Phi_B$ modify the energy eigenvalues where, the degeneracy of the energy levels is broken by the torsion parameter. 

We have seen in each cases that the angular quantum number $l$ is shifted, $l \rightarrow l_{eff}=\frac{1}{\alpha}\,(l-\Phi-k\,\chi)$, an effective angular quantum number. Thus, the relativistic energy eigenvalues obtained in {\it sub-section 2.1--2.2} depends on the geometric quantum phase \cite{YA,GAM} as well as the torsion parameter and only the geometric quantum phase in {\it sub-section 2.3}. This dependence of the relativistic energy eigenvalues on the geometric quantum phase gives rise to a relativistic analogue of the Aharonov-Bohm effect for bound states \cite{FA,EVBL3,RLLV5,MP,VBB,CF2,CF5,EVBL,aa12,EVBL2,EPaa12}. Besides, we have that $E_{n, l} (\Phi_B+\Phi_0)=E_{n, l\mp \tau} (\Phi_B)$ where, $\Phi_0=\pm\,\frac{2\,\pi}{e}\,\tau$ with $\tau=0,1,2...$.

It is well known in non-relativistic quantum mechanics that the Landau quantization is the simplest system that would work with the studies of quantum Hall effect. Therefore, the relativistic quantum systems analyzed in this work would used for investigating the influence of torsion, the cosmic string as well as the Cornell-type potential on the thermodynamic properties of quantum systems \cite{HHa,ME,ANI,HHa2}, searching a relativistic analogue of the quantum Hall effect \cite{REP,BB}, and the displaced Fock states \cite{JL} in a topological defects space-time with a spacelike dislocation.

\section*{Data Availability}

No data has been used to prepare this paper.

\section*{Conflict of Interest}

Author declares that there is no conflict of interest regarding publication this paper.

\section*{Acknowledgment}

Author sincerely acknowledge the anonymous kind referee(s) for their valuable comments and suggestion which have greatly improved the present paper.


\begin{thebibliography}{99}

\bibitem{aa1} B. D. B. Figueiredo, I. D. Soares and J. Tiomno, Class. Quantum Grav. {\bf 9}, 1593 (1992).

\bibitem{aa2} F. Ahmed, Eur. Phys. J. C (2018) {\bf 78} : 598.

\bibitem{aa3} Z. Wang, Z. Long, C. Long and M. Wu, Eur. Phys. J. Plus (2015) {\bf 130} : 36.

\bibitem{aa4} J. Carvalho, A. M. de M. Carvalho and C. Furtado, Eur. Phys. J. C (2014) {\bf 74} : 2935.

\bibitem{aa5} R. L. L. Vit\'{o}ria, C. Furtado and K. Bakke, Eur. Phys. J. C (2018) {\bf 78} : 44.

\bibitem{aa6} F. Ahmed, Eur. Phys. J. C (2019) {\bf 79} : 682.

\bibitem{aa7} F. Ahmed, Eur. Phys. J. C (2019) {\bf 79} : 104.

\bibitem{aa8} N. Drukker, B. Fiol and J. Simon, JCAP (2004) {\bf 0410} : 012.

\bibitem{aa9} S. Das and J. Gegenberg, Gen. Relativ. Grav. (2008) {\bf 40} : 2115.

\bibitem{aa10} C. Furtado, B. C. G. da Cunha, F. Moraes, E. R. B. de Mello and V. B. Bezerra, Phys. Lett. {\bf A 195}, 90 (1994).

\bibitem{aa12} J. Carvalho, A. M. M. Carvalho and E. Cavalcante and C. Furtado, Eur. Phys. J. C (2016) {\bf 76} : 365.

\bibitem{aa15} F. Ahmed, Eur. Phys. J. Plus (2020) {\bf 135} : 108.

\bibitem{HS} H. Sobhani, H. Hassanabdi and W. S. Chung, Int. J. Geom. Meths. Mod. Phys. {\bf 15}, 1850037 (2018).

\bibitem{AOP} F. Ahmed, Ann. Phys. (N. Y.) {\bf 401}, 193 (2019).

\bibitem{AOP2} F. Ahmed, Ann. Phys. (N. Y.) {\bf 404}, 1 (2019).

\bibitem{GERG} F. Ahmed, Gen. Relativ. Grav. (2019) {\bf 51} : 69.

\bibitem{AOP3} F. Ahmed, Ann. Phys. (N. Y.) {\bf 411}, 167941 (2019).

\bibitem{GERG2} F. Ahmed, Gen. Relativ. Grav. (2019) {\bf 51} : 129.

\bibitem{ERFM} E. R. F. Medeiros and E. R. Bezerra de Mello, Eur. Phys. J. C (2012) {\bf 72} : 2051.

\bibitem{EPaa123} F. Ahmed, Eur. Phys. J. C (2019) {\bf 79} : 534.

\bibitem{MdM2} M. de Montigny, S. Zare and H. Hassanabadi, Gen. Relativ. Grav. (2018) {\bf 50} : 47.

\bibitem{EPaa122} P. Sedaghatnia, H. Hassanabadi and F. Ahmed, Eur. Phys. J. C (2019) {\bf 79} : 541.

\bibitem{EPJP2} F. Ahmed, Eur. Phys. J. Plus (2019) {\bf 134} : 518; {\it ibid} 582.

\bibitem{AV3} A. Vilenkin and E. P. S. Shellard, {\it Cosmic strings and other topological defects}, Cambridge University Press, Cambridge (1994).

\bibitem{MH} M. B. Hindmarsh and T. Kibble, Rep. Prog. Phys. {\bf 58}, 477 (1995).

\bibitem{AV} A. Vilenkin, Phys. Lett. {\bf B 133}, 177 (1983).

\bibitem{AV2} A. Vilenkin, Phys. Rep. {\bf 121}, 263 (1985).

\bibitem{WAH} W. A. Hiscock, Phys. Rev. {\bf A 31}, 3288 (1985).

\bibitem{BL} B. Linet, Gen. Relativ. Grav. {\bf 17}, 1109 (1985).

\bibitem{MOK} M. O. Katanaev and I. V. Volovich, Ann. Phys. (N. Y.) {\bf 216}, 1 (1992).

\bibitem{CF} C. Furtado and F. Moraes, Phys. Lett. {\bf A 188}, 394 (1994).

\bibitem{TWBK} T. W. B. Kibble, J. Phys. {\bf A 19}, 1387 (1976).

\bibitem{LCNS2} L. C. N. Santos and C. C. Barros Jr., Eur. Phys. J. C (2018) {\bf 78} : 13.

\bibitem{H1} M. Hosseini, H. Hassanabadi, S. Hassanabadi and P. Sedaghatnia, Int. J. Geom. Meths. Mod. Phys. {\bf 16}, 1950054 (2019).

\bibitem{H2} M. Hosseinpour, H. Hassanabadi and F. M. Andrade, Eur. Phys. J. C (2018) {\bf 78} : 93.

\bibitem{H4}  M. Hosseinpour, F. M. Andrade, E. O. Silva and H. Hassanabadi, Eur. Phys. J. C (2017) {\bf 77} : 270.

\bibitem{H5} H. Sobhani, H. Hassanabdi and W. S. Chung, Eur. Phys. J. C (2018) {\bf 78} : 106.

\bibitem{MS}  M. Hosseinpour and H. Hassanabadi, Int. J. Mod. Phys. A {\bf 30}, 1550124 (2015). 

\bibitem{MdM} M. de Montigny, M. Hosseinpour, H. Hassanabadi, Int. J. Mod. Phys. A {\bf 31}, 1650191 (2016).

\bibitem{bb1} S. Zare, H. Hassanabadi and M de. Montigny, Gen. Relativ. Grav. (2020) {\bf 52} : 1.

\bibitem{bb2} M. Hosseinpour, H. Hassanabadi and M. de Montigny, Eur. Phys. J. C (2019) {\bf 79} : 311.

\bibitem{bb4} H. Hassanabadi, S. zare and M. de Montigny, Gen. Relativ. Grav. (2018) {\bf 50} : 104.

\bibitem{bb6} M. Hosseinpour and H. Hassanabadi, Adv. High. Energy Phys. {\bf 2018}, 2959354 (2018). 

\bibitem{bb7} H. Hassanadabi, W. C. Chung, S. Zare and H. Sobhani, Eur. Phys. J. C (2018) {\bf 78} : 83.

\bibitem{bb8} M. Hosseinpour and H. Hassanabadi, Eur. Phys. J. Plus (2015) {\bf 130} : 236.

\bibitem{bb9} H. Hassanadabi, A. Afshadoost and S. Zarrinkamar, Ann. Phys. (N. Y.) {\bf 356}, 346 (2015).

\bibitem{bb10}  A. Afshadoost and H. Hassanadabi, Can. J. Phys. {\bf 94}, 71 (2015).

\bibitem{MKB} M. K. Bahar and F. Yasuk, Adv. High Energy Phys. {\bf 2013}, 814985 (2013).

\bibitem{CA} C. Alexandrou, P. de Forcrand and O. Jahn, Nuc. Phys. B (Proc. Suppl.) {\bf 119}, 667 (2003).

\bibitem{MH2} M. Hosseinpour, H. Hassanabadi and M. de Montigny, Int. J. Geom. Meths. Mod. Phys. {\bf 15}, 1850165 (2018).

\bibitem{MPLA} F. Ahmed, Mod. Phys. Lett. {\bf A 34}, 1950314 (2019).

\bibitem{FA} F. Ahmed, Adv. High Energy Phys. {\bf 2020}, 5691025 (2020).

\bibitem{RLLV3} R. L. L. Vit\'{o}ria and K. Bakke, Eur. Phys. J. Plus (2016) {\bf 131} : 36.

\bibitem{EVBL3} E. V. B. Leite, H. Belich and R. L. L. Vit\'{o}ria, Adv. High Energy Phys. {\bf 2019}, 6740360 (2019).

\bibitem{MSC}  M. S. Cunha, C. R. Muniz, H. R. Christiansen and V. B. Bezerra, Eur. Phys. J. C (2016) {\bf 76} : 512.

\bibitem{RLLV4}  R. L. L. Vit\'{o}ria and H. Belich, Eur. Phys. J. C (2018) {\bf 78} : 999.

\bibitem{RLLV2}  R. L. L. Vit\'{o}ria and H. Belich, Adv. High Energy Phys. {\bf 2019}, 1248393 (2019).

\bibitem{ALCO} A. L. C. de Oliveira and E. R. Bezerra de Mello, Class. Quantum Grav. {\bf 23}, 5249 (2006).

\bibitem{RLLV5} R. L. L. Vit\'{o}ria and K. Bakke, Int. J. Mod. Phys. {\bf D 27}, 1850005 (2018).

\bibitem{RLLV} R. L. L. Vit\'{o}ria and K. Bakke, Eur. Phys. J. Plus (2018) {\bf 133} : 490.

\bibitem{MP} M. Peshkin and A. Tonomura, {\sf The Aharonov-Bohm effect}, {\it Lect. Notes Phys.} {\bf Vol. 340}, Springer, Berlin, Germany (1989).

\bibitem{VBB} V. B. Bezerra, J. Math. Phys. {\bf 30}, 2895 (1989).

\bibitem{RAP} R. A. Puntigam and H. H. Soleng, Class. Quantum Grav. {\bf 14}, 1129 (1997).

\bibitem{VBB10} V. B. Bezzera, J. Math. Phys. {\bf 38}, 2553 (1997).

\bibitem{HK} H. Kleinert, {\it Gauge Fields in Condensed Matter}, {\bf Vol. 2}, World Scientific Publishing Co, Singapore (1989).

\bibitem{CF3} C. Furtado and F. Moraes, Euro. Phys. Lett. {\bf 45}, 279 (1999).

\bibitem{JW} J. Wang, K. Ma, K. Li and H. Fan, Ann. Phys. (N. Y.) {\bf 362}, 327 (2015).

\bibitem{CFil} C. Filgueiras, M. Rojas, G. Aciole and E. O. Silva, Phys. Lett. {\bf A 380}, 3847 (2016).

\bibitem{GAM2} G. A. Marques, V. B. Bezerra, C. Furtado and F. Moraes, Int. J. Mod. Phys. {\bf A 20}, 6051 (2005).

\bibitem{CF4} C. Furtado, V. B. Bezerra and F. Moraes, Phys. Lett. A {\bf 289}, 160 (2001).

\bibitem{RLLV6} R. L. L. Vit\'{o}ria, Eur. Phys. J. C (2019) {\bf 79} : 844.

\bibitem{RLLV8} R. L. L. Vit\'{o}ria and K. Bakke, Eur. Phys. J. C (2018) {\bf 78} : 175.

\bibitem{LCNS} L. C. N. Santos and C. C. Barros, Eur. Phys. J. C (2017) {\bf 77} : 186.

\bibitem{RLLV7} R. L. L. Vit\'{o}ria and K. Bakke, Gen. Relativ. Grav. (2016) {\bf 48} : 161.

\bibitem{YA} Y. Aharonov and D. Bohm, Phys. Rev. {\bf 115}, 485 (1959).

\bibitem{GAM} G. de A. Marques, C. Furtado, V. B. Bezerra and F. Moraes, J. Phys. A : Math. Theor. {\bf 34}, 5945 (2001).

\bibitem{CF6} C. Furtado and F. Moraes, J. Phys. A : Math. Theor. {\bf 33}, 5513 (2000).

\bibitem{RJ} R. Jackiw, A. I. Milstein, S.-Y. Pi and I. S. Terekhov, Phys. Rev. {\bf B 80}, 033413 (2009).

\bibitem{MAA} M. A. Anacleto, I. G. Salako, F. A. Brito and E. Passos, Phys. Rev. {\bf D 92}, 125010 (2015).

\bibitem{VRK} V. R. Khalilov, Eur. Phys. J. C (2014) {\bf 74} : 2708.

\bibitem{CC} C. Coste, F. Lund and M. Umeki, Phys. Rev. {\bf E 60}, 4908 (1999).

\bibitem{CF2} C. Furtado, V. B. Bezerra and F. Moraes, Mod. Phys. Lett. {\bf A 15}, 253 (2000).

\bibitem{CF5} C. Furtado, F. Moraes and V. Bezerra, Phys. Rev. {\bf D 59}, 107504 (1999).

\bibitem{EVBL} E. V. B. Leite, K. Bakke and H. Belich, Adv. High Energy Phys. {\bf 2015}, 925846 (2015).

\bibitem{EVBL2} E. V. B. Leite, R. L. L. Vit\'{o}ria and H. Belich, Mod. Phys. Lett. {\bf A 34}, 1950319 (2019).

\bibitem{EPaa12} F. Ahmed, Eur. Phys. J. C (2020) {\bf 80} : 211.

\bibitem{HB} H. Belich, E. O. Silva, M. M. Ferreira Jr. and M. T. D. Orlando, Phys. Rev. {\bf D 83}, 125025 (2011).

\bibitem{WG} W. Greiner, {\it Relativistic Quantum Mechanics : Wave Equations}, 3rd edition, Springer-Verlag, Berlin, Germany (2000).

\bibitem{AR} A. Ronveaux, {\it Heun’s Differential Equations}, Oxford University Press, Oxford ( 1995).

\bibitem{SYS} S. Y. Slavyanov and W. Lay, {\it Special Functions: A Unified Theory Based in Singularities},  Oxford 
University Press, New York (2000).

\bibitem{GBA} G. B. Arfken and H. J. Weber, {\it Mathematical Methods For Physicists}, Elsevier Academic Pres, London (2005).

\bibitem{AVV} A Vercin, Phys. Lett. B {\bf 260}, 120 (1991).

\bibitem{JM} J. Myrhein, E. Halvorsen and A. Vercin, Phys. Lett. B {\bf 278}, 171 (1992).

\bibitem{HHa} H. Hassanabadi and M. Hosseinpour, Eur. Phys. J. C (2016) {\bf 76} : 553. 

\bibitem{ME} M. Eshghi and H. Mehraban, Eur. Phys. J. Plus (2017) {\bf 132} : 121.

\bibitem{ANI} A. N. Ikot, B. C. Lutfuoglu, M. I. Ngwueke, M. E. Udoh, S. Zare and H. Hassanabadi, Eur. Phys. J. Plus (2016) {\bf 131} : 419.

\bibitem{HHa2} H. Hassanabadi, S. Sargolzaeipor and B. H. Yazarloo, Few-Body System {\bf 56}, 115 (2015).

\bibitem{REP} R. E. Prange and S. M. Girvin, {\it The Quantum Hall Effect}, Springer-Verlag, New York (1990).

\bibitem{BB} B. Basu and D. Chowdhury, Phys. Rev. {\bf D 90}, 125014 (2014).

\bibitem{JL} J. L. de Melo, K. Bakke and C. Furtado, Eur. Phys. Lett. {\bf 115}, 20001 (2016).

\end{thebibliography}
\end{document}